# Self-imaging silicon Raman amplifier

**Varun Raghunathan[1], Hagen Renner[2], Robert R. Rice[3] and Bahram Jalali[1]**

[1] *Electrical Engineering Department, UCLA, 420 Westwood Plaza, Los Angeles, CA 90095-1594, USA;*
[2] *Technische Universität Hamburg-Harburg, 21071 Hamburg, Germany*
[3] *Northrop Grumman Space Technology, One Space Park, Redondo Beach, CA 90278, USA*
*jalali@ucla.edu*

**Abstract:** We propose a new type of waveguide optical amplifier. The device consists of collinearly propagating pump and amplified Stokes beams with periodic imaging of the Stokes beam due to the Talbot effect. The application of this device as an Image preamplifier for Mid Wave Infrared (MWIR) remote sensing is discussed and its performance is described. Silicon is the preferred material for this application in MWIR due to its excellent transmission properties, high thermal conductivity, high damage threshold and the mature fabrication technology. In these devices, the Raman amplification process also includes four-wave-mixing between various spatial modes of pump and Stokes signals. This phenomenon is unique to nonlinear interactions in multimode waveguides and places a limit on the maximum achievable gain, beyond which the image begins to distort. Another source of image distortion is the preferential amplification of Stokes modes that have the highest overlap with the pump. These effects introduce a tradeoff between the gain and image quality. We show that a possible solution to this trade-off is to restrict the pump into a single higher order waveguide mode.

**1. Introduction**

The stimulated Raman scattering effect has been used successfully to realize amplifiers and lasers in various solid state media including optical fibers [1], semiconductors such as GaP [2], silicon [3] and more exotic solid state media such as, $Ba(NO_3)_2$, $LiIO_3$, and $KGd(WO_4)_2$ etc [4]. The Raman process in general can achieve high gains by using suitable interaction lengths and modal area without the need for complicated phase-matching schemes as in the case of Optical Parametric Oscillator (OPO) and Optical parametric amplifier (OPA). However, for high power applications, direct power-scaling of Raman amplifiers by increasing the pump power, keeping the modal area constant, is challenged by: (i) the lack of high power pump sources with good beam quality, and (ii) by the possibility of damaging the medium at high optical intensities. Problems also exists in bulk media where increasing the pump power results in beam distortion due to thermal lensing and self focusing. These problems can be addressed by concomitant scaling of modal area with the increase in pump power. The scaling of pump power / modal area invariably results in multiple spatial modes in waveguide structures. The Raman interactions in the presence of multiple spatial modes were first studied by Bloembergen et al. [5-7], who described interesting effects such as increased Raman gain in an amplifier due to interaction of higher order pump and Stokes modes, and the so-called Raman beam clean-up effect. More recently the beam clean-up effect has been observed in bulk Raman crystals [4] and in multimode optical fibers [8].

Certain multimode waveguide structures also exhibit Talbot self-imaging effect on account of constructive interference among the various waveguide modes every periodic length [9]. This phenomenon also takes place when the multimode waveguide is comprised of an optically amplifying medium. In such multimode waveguides with an active gain medium, the input electric field distribution is amplified and replicated at the focal points, which correspond to full Talbot planes. These amplifier configurations offer the benefits of good beam quality at the focal points along with better light-gain medium interaction and also reduction of deleterious effects such as self-focusing and thermal lensing. So far, the use of amplifiers with Talbot imaging has been restricted to rare-earth doped solid state medium, with top or side pumping using diodes [10,11].

In this paper we propose a multimode silicon waveguide Raman amplifier that consists of collinearly propagating mid infrared pump and Stokes beams. The waveguide amplifies and images the spatial profile of an input beam. We describe a coupled-mode analysis that includes the conventional Raman amplification, and the four-wave mixing that occurs between spatial pump and Stokes modes, due to the Raman nonlinearity. We find that the conventional Raman amplification term and the Raman Spatial FWM (RS-FWM) start to distort the amplified image beyond a certain waveguide length or pump power. The prospects of using this device as an image preamplifier in MWIR Laser Radar (LADAR) are discussed. Image amplifiers in the near-infrared region (~1-1.5μm) have been implemented in the past using rare earth doped fibers or other solid state media [12,13]. The recent progress made in the MWIR laser sources and the richness of the spectral signatures of molecules in these wavelengths has opened up new applications of MWIR sources in LADAR and standoff chemical detection [14]. The range and sensitivity of these remote sensing applications can be improved by the use of an optical preamplifier before signal detection.

The requirements imposed on a gain medium for the image amplifier includes low linear and nonlinear optical loss at both the pump and Stokes wavelengths, high optical damage threshold, high Raman gain coefficient, high thermal conductivity and the availability of large samples with high crystal quality. A material that meets all these criteria in the MWIR is single crystal silicon. In all respects, silicon is an excellent choice for the high power MWIR applications, especially since the problem of two photon excitation and free carrier absorption are eliminated when the pump wavelength is longer than the two-photon absorption edge of 2.3 μm [15]. Such a technology will also benefit from the mature silicon fabrication processes. Moreover, the device offers large field-of-view image amplification due to the high numerical aperture of the multimode Silicon waveguide.

## 2. Modeling of a multimode Raman amplifier

Multimode interference in passive waveguides have applications in optical splitters and couplers, a detailed account of which can be found in ref [9]. The multimode silicon waveguides analyzed in this work consist of a silicon thin film structure as shown in figure 1. The waveguide width, $a = 80$μm, and its thickness $b=50$μm. Under symmetric (on-axis) launch with a 40μm beam, the optical power is coupled mainly (~98%) in the fundamental mode in the Y-direction and the multimode nature of the waveguide considered in the X-direction. This simplifies the numerical solutions and helps to elucidate the key features of the device, without loss of generality. The analysis can be naturally extended to other waveguide dimensions.

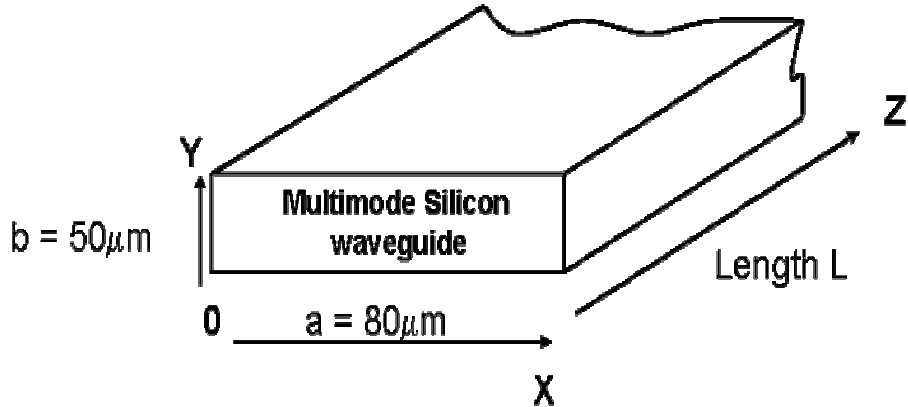

Fig.1. Cross-section of multimode silicon waveguides used in this analysis. For the ease of analysis, waveguide width, $a$ is taken to be much larger than the thickness, $b$.

The orthonormal eigen-modes of the 2-D planar waveguide are taken to be of sinusoidal form:

$$\phi_{mn} = \sqrt{\frac{4Z}{ab}} Sin\left(\frac{m\pi x}{a}\right) Sin\left(\frac{n\pi y}{b}\right), \qquad 0 < x < a \text{ and } 0 < y < b \qquad (1)$$

with mode index $(m,n)$ being integers $\geq 1$ in general. Z is the impedance of the medium. This mode profile assumes that there is no evanescent tail of the mode present in the cladding region, which is approximately true for large core (multimode) high-index contrast silicon waveguides.

For ease of analysis, input pump and Stokes beams are assumed to be plane-waves with transverse Gaussian profile as shown below:

$$\psi_{in} = \frac{\sqrt{PZ}}{w\sqrt{2\pi}} \cdot \exp\left(\frac{-(x-a/2)^2}{4w^2}\right) \exp\left(\frac{-(y-b/2)^2}{4w^2}\right) \cdot \exp(j\theta) \tag{2}$$

$P$ refers to the laser power, the beams are assumed to be launched with $1/e^2$ beam radius of $2w$ and launched centered with respect to X and Y axis ($a/2$ and $b/2$ respectively). The input phase factor is taken as $\theta$.

The evolution of the optical field along the waveguide is written as:

$$\psi(z) = \sum_m \sum_n A_{mn}(z)\phi_{mn} e^{j\beta_{mn} z} \tag{3}$$

where $A_{mn}$ and $\beta_{mn}$ are the mode coefficient and propagation constant for the mode ($m,n$). The multimode imaging phenomenon arises due to the periodic constructive interference of the modes in the waveguide at the *focal points*, where the input profile is essentially reproduced. The focal points occurs periodically at intervals [9]:

$$L_{image} \approx p\frac{4n_0 a^2}{\lambda_0} \tag{4}$$

where, even and odd values of $p$ lead to original and reversed images of the input beam respectively. The self-imaging points which reproduce the original image are referred to as focal points in this paper. For the passive device (ie. no gain), the mode coefficients $A_{mn}$ are essentially constant along the waveguide. However, in the case of the active device (ie. for a Raman amplifier) to be considered next, the evolution of the mode coefficients along the waveguide are also to be accounted for.

Table 1 lists the parameters used in the simulations. It was found that for the launch condition considered, 11 modes along the X-direction and 1 mode along the Y-direction accounted for ~ 98% of the coupled optical power and was sufficient for the purpose of this simulation. Based on the parameter as listed in Table I, imaging length at the pump and Stokes wavelengths computed using Eq. (4) are ~ 6cm and 5.07cm respectively. The imaging lengths are slightly different due to chromatic dispersion.

**Table 1.** List of parameters used throughout this paper and the values of these parameters used in these simulations

| Parameters | Value |
|---|---|
| Input beam profile | $1/e^2$ diameter of the gaussian beam = 40µm |
| Waveguide dimensions | a = 80µm, b = 50µm |
| Simulation grid size | $\Delta x$ = 1µm, $\Delta y$ = 2µm, $\Delta z$ =100µm |
| Waveguide modes | X-direction: 11<br>Y-direction: 1 |
| n – (Sellemier model)<br>($\lambda_g$ bandgap in µm) | $n^2 = 11.6858 + \frac{0.939816}{\lambda^2} + \frac{8.1 \times 10^{-3} \times \lambda_g^2}{\lambda^2 - \lambda_g^2}$ |
| Wavelengths | Pump: 2.936µm<br>Stokes: 3.466µm |
| Raman susceptibility | $1.6 \times 10^{-18}$ m$^2$/V$^2$ |
| Electronic susceptibility | $0.5 \times 10^{-18}$ m$^2$/V$^2$ |

In the case of the active multimode waveguide with Raman gain, the evolution of the pump and Stokes mode coefficients are described by a coupled-mode analysis as follows:

$$\frac{dA_{S-mn}}{dz} = -\frac{\alpha_{S-mn}}{2} A_{S-mn} + \sum_{kl} \kappa_{mn-kl} |A_{P-kl}|^2 A_{S-mn} + \sum_{\substack{k \neq m \& \\ l \neq n}} \kappa'_{mn-kl} A_{P-mn} A_{P-kl}^* A_{S-kl} e^{j\Delta\beta z} \quad (5.1)$$

$$\frac{dA_{P-mn}}{dz} = -\frac{\alpha_{P-mn}}{2} A_{P-mn} - \frac{\omega_P}{\omega_S} \sum_{kl} \kappa_{mn-kl} |A_{S-kl}|^2 A_{P-mn} - \frac{\omega_P}{\omega_S} \sum_{\substack{k \neq m \& \\ l \neq n}} \kappa'_{mn-kl} A_{S-mn} A_{S-kl}^* A_{P-kl} e^{-j\Delta\beta z} \quad (5.2)$$

$$\kappa_{mn-kl} = \omega_S \varepsilon_o \int_0^b \int_0^a \phi_{P-kl} \phi_{P-kl}^* (\chi_{Raman}^{(3)}) \phi_{S-mn} \phi_{S-mn}^* \, dx dy \quad (5.3)$$

$$\kappa'_{mn-kl} = \omega_S \varepsilon_o \int_0^b \int_0^a \phi_{P-kl}^* \phi_{P-mn} (\chi_{Raman}^{(3)}) \phi_{S-mn}^* \phi_{S-kl} \, dx dy \quad (5.4)$$

$$\Delta\beta = (\beta_{P-mn} - \beta_{P-kl}) - (\beta_{S-mn} - \beta_{S-kl}) \quad (5.5)$$

This set of coupled-mode equations for a multimode waveguide can be considered as an extension of the two-mode treatment presented in ref. [16] for the case of optical fibers. Here, $A_{S-mn}$ and $A_{P-mn}$ refer to the Stokes and pump mode coefficients respectively for mode index (m,n), $\phi_{S-mn}$ and $\phi_{P-mn}$ refer to the mode profiles of the Stokes and pump as given by Eq. (1), $\alpha_{S-mn}$ and $\alpha_{P-mn}$ refer to the linear propagation losses of the modes, which in general can be different for different modes, $\kappa_{mn-kl}$ and $\kappa'_{mn-kl}$ refer to the coupling coefficients of the Raman interaction process and $\Delta\beta$ refers to the phase mismatch for the Raman Spatial FWM (RS-FWM) process which is described below. We have ignored the RS-FWM terms which are highly oscillatory as this behavior is expected to wash away its contribution. Throughout this analysis we assume that the waveguide eigenmodes do not change as a result of the nonlinear polarization effects. This is valid as $\chi^{(3)}$ is ~ $10^{18}$ times weaker than the linear $\chi^{(1)}$ and hence the third order polarization $P^{(3)}$ is much weaker than $P^{(1)}$. Thus the only effect of the nonlinear interaction is to alter the amplitude and phase of the mode coefficients. The Raman gain experienced by the Stokes modes is represented by the coupling coefficient, $\kappa_{mn-kl}$ and $\kappa'_{mn-kl}$, which are functions of the Raman susceptibility and the extent of overlap among the interacting modes. Eqs (5.1) and (5.2) do not include self phase and cross phase modulation effects. This assumption is valid because for the input intensity levels considered here (up to 25 MW/cm$^2$), the nonlinear length, $L_{NL} = \frac{cA_{eff}}{n_2 \omega_0 P}$ [1], is ~13cm, which is larger than the typical waveguide length considered in this analysis (~5cm). Any effect of self-phase and cross-phase modulation becomes pronounced only at lengths comparable to the nonlinear length.

There are two types of Raman interactions which occur in a multimode waveguide. The first term on the right hand side of Eq. (5.1) is the conventional Raman amplification of the Stokes signal. The corresponding gain is independent of the pump phase as it is only proportional to the pump intensity, $|A_{P-kl}|^2$.

The second terms on the right hand side of Eq. (5.1) describes RS-FWM terms which arise due to the mixing of the different spatial modes of pump and Stokes signals. This term has the expected dependence of FWM on phase mismatch (Eq. (5.5)). If phase matching can be ensured this term will contribute to the overall gain and will be beneficial as it leads to gain enhancement. The gain enhancement was predicted by Bloembergen [5] and has been observed in the form of reduced pump threshold for stimulated Raman scattering in a gas cell [6,7].

In the context of image amplification, there are two mechanisms that can affect the desired self-imaging of the amplified Stokes beam. Both lead to a trade-off between the achievable gain and the reproducibility and quality of the Stokes image at focal points. First, with increasing length and pump power the Stokes modes which have highest overlap with the pump modes experience higher gains compared to the other Stokes modes. This leads to preferential amplification of certain Stokes modes and hence distortion of the image. Secondly, the RS-FWM effect changes the phases of Stokes mode coefficients and further affects the image. The RS-FWM effect has another implication as can be seen in the $2^{nd}$ term on the right side of Eq. (5.1), when multiple free running pump lasers are used to provide high pump powers, the transfer of random pump phases to the amplified Stokes modes can also degrade the image quality.

For MWIR applications considered here, the pump and Stokes photons are below the two-photon bandedge, hence two-photon absorption and the concomitant free-carrier losses are nonexistent [15]. Silicon being indirect bandgap is expected to have insignificant three-photon absorption when compared to other shorter bandgap semiconductors such as GaAs [18].

The coupled mode equations presented above were solved numerically using a finite difference algorithm. The simulator grid size in the x, y and z directions are taken to be 1μm, 2μm and 100μm respectively. For the purpose of studying the Raman amplification process, the input pump was considered to be 1KW peak power or 1mW average power at the wavelength of 2.936μm. This may be achievable in solid-state laser systems under quasi-CW conditions at energy levels of ~100μJ over 100nsec pulse [19]. The input peak intensity of the pump is ~25MW/cm$^2$ and is comparable to the intensities used in the near infrared to study Raman scattering in silicon. The multimode Raman amplifier presented in this work is power scalable and the Raman amplification process can be made to work at lower or higher power levels through appropriate waveguide dimension scaling. This would also change the imaging length as given by Eq. (4). The input Stokes beam is assumed to be at 1μW at 3.466μm wavelength. Figure 2 shows the electric-field contour (X-Z profile) of the pump and Stokes as they propagate along the waveguide. The amplification of the Stokes is clearly noticeable along with the self-imaging effect. The pump is not significantly depleted due to the weak input Stokes beam. It is found that the pump and Stokes fields self-image much more frequently than the focal length as calculated using eq. (4). This is due to the fact that we have considered a symmetrical input field in this example. In this case, the image will reproduce itself 8 times within the focal length of a general input field, $L_{image}$, described by Eq. (4) [9].

Figure 3 shows the small signal Raman gain that can be achieved along the waveguide for varying waveguide propagation losses. The vertical dashed lines denote the location of the $1^{st}$ and $2^{nd}$ focal points. For the case of zero loss, Raman gain of 10dB is reached at the first focal point for peak pump power levels of ~1KW. For such large area silicon waveguide, losses are expected to be on the order of 0.1dB/cm [17]. As seen in this figure, for the propagation losses considered here, 0.1dB/cm, 0.2dB/cm and 0.5dB/cm, the gain at the first imaging length are 9dB, 8dB, and 5dB, respectively.

The impact of linear losses need to be explained. If the linear losses increase with the mode number then a spatial low-pass filtering will occur as the amplitudes of the higher modes are diminished, while their phases remain unaffected. This would be the case if losses occur primarily on the waveguide surfaces. In addition, propagation losses also impact the image quality by diminishing the contribution of the RS-FWM effect as the image propagates along the waveguide.

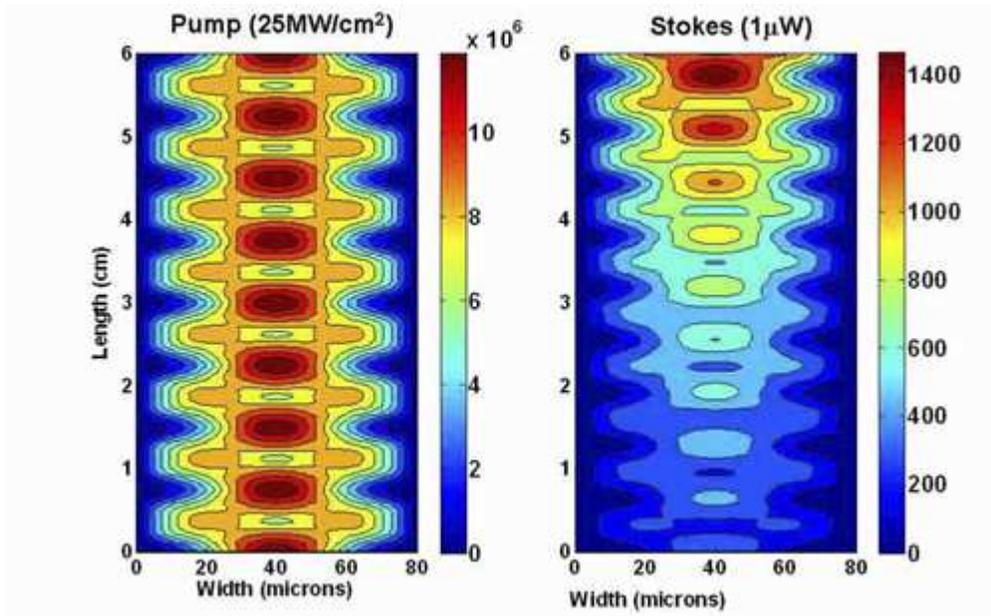

Fig. 2. Contour profile of the electric field amplitude (X-Z profile) showing the self-imaging Raman amplifier with the evolution of the pump and Stokes along the length of the multimode silicon waveguide. A single pump and Stokes Gaussian beam is launched into the waveguide. Pump power coupled into the waveguide is 1KW peak (1mW average) and Stokes power is 1µW.

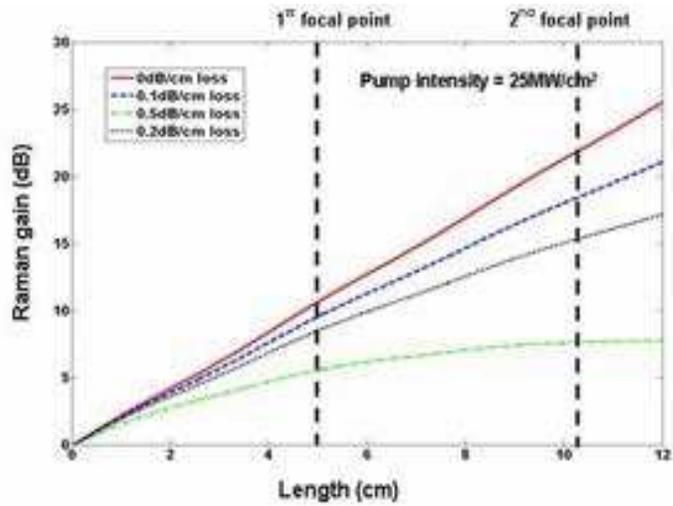

Fig. 3. Evolution of small-signal Raman gain along the length of the multimode silicon Raman amplifier. The input pump and Stokes power are taken to be 1KW peak (1mW average) and 1µW respectively.

To assess the impact of Raman interaction on the self imaging phenomenon, we have computed the well known $M^2$ parameter for the beam profile, and its evolution along the waveguide. Following the known procedure [20], the $M^2$ parameter was computed by taking the Fourier transform of the transversal beam profile. The $M^2$ parameter indicates deviations from a Gaussian profile with $M^2=1$ corresponding to an ideal Gaussian beam, such as the input profile considered in the present case. Figure 4 shows the $M^2$ parameter for the Stokes beams for the following cases: (a) Stokes beam propagating in a passive waveguide without Raman amplification (no pump) and (b) in the presence of Raman amplification with input pump intensities of 25 and 50 MW/cm$^2$. The input beam is found to reproduce itself periodically along the waveguide with $M^2$ reaching it minima at focal points. In the absence of Raman interactions (Figure 4(a)) self imaging is perfect with $M^2=1$ reproduced at all focal points. In the presence of Raman amplification however, the image begins to be distorted while it is amplified along the waveguide (figure 4(b)). This distortion in image quality is found to be more significant with increase in pump power and waveguide length. The slight increase in $M^2$ at focal points (leading to loss of image quality) and the overall reduction in its average value are due to the preferential amplification of the fundamental mode. The periodicity of $M^2$ along the waveguide length is also modulated by ripples due to the RS-FWM terms in Eqs. (5.1) and (5.2) by contributing addition phase factors to the mode coefficients. Thus, it is clear that there exists a tradeoff between the amount of gain that can be achieved and the image distortion. From figures 3 and 4 it is found that gains close to 10dB can be achieved in waveguide lengths of ~5cm with minimal image distortion.

**3. Image Amplification**

The range and sensitivity of LADAR (Laser Detection And Ranging) and remote sensing systems can be improved by the use of optical preamplifiers before detection. So far, efforts to implement optical image preamplifiers have been mainly focused on near infrared LADAR systems using doped fiber, YAG-based amplifiers and Barium Nitrate Raman medium [12, 21, 13]. There is increasing interest in LADAR systems operating in the MWIR due to strong vibrational resonance of molecules in this range and the recent availability of efficient MWIR laser sources [14]. Moreover, the solid-state Raman media used previously for image amplification consisted of a bulk crystal with no waveguiding [13]. The lack of waveguiding, and hence self imaging, limits the interaction length before image quality is deteriorated. The waveguide implementation introduced here increases the length over which image quality can be maintained and also reduces the threshold pump power through confinement of the pump mode. The use of silicon as the active medium also provides many advantages such as a high Raman coefficient, high thermal conductivity, high optical damage threshold, and mature fabrication technology.

To assess the performance of the device as an image amplifier, simple test images were launched into the device input and its evolution through the waveguide was computed using the methodology described above. Fig. 5 shows the cross-sectional amplitude profiles at the input and output at the first imaging length. For the purpose of this simulation we have considered 20 waveguide modes along the X-direction and 1 mode along the Y-direction. We emphasize the transversal profile is shown here as opposed to the longitudinal view shown in Fig. 2. A pump beam with 1KW peak (1mW average) power is assumed to be launched centered with respect to the waveguide. The periodic patterns considered here are 50, 100 and 200 lines per mm ruling. The center portion of the image appears brighter than the edges due to the preferential amplification of the fundamental Stokes modes. The deterioration is more severe when the image contains higher spatial frequencies. Nonetheless the rulings are clearly resolvable at amplification level (~10dB gain).

The preferential amplification of the fundamental Stokes mode leading to loss of image quality can be eliminated using selective pump mode excitation as described below.

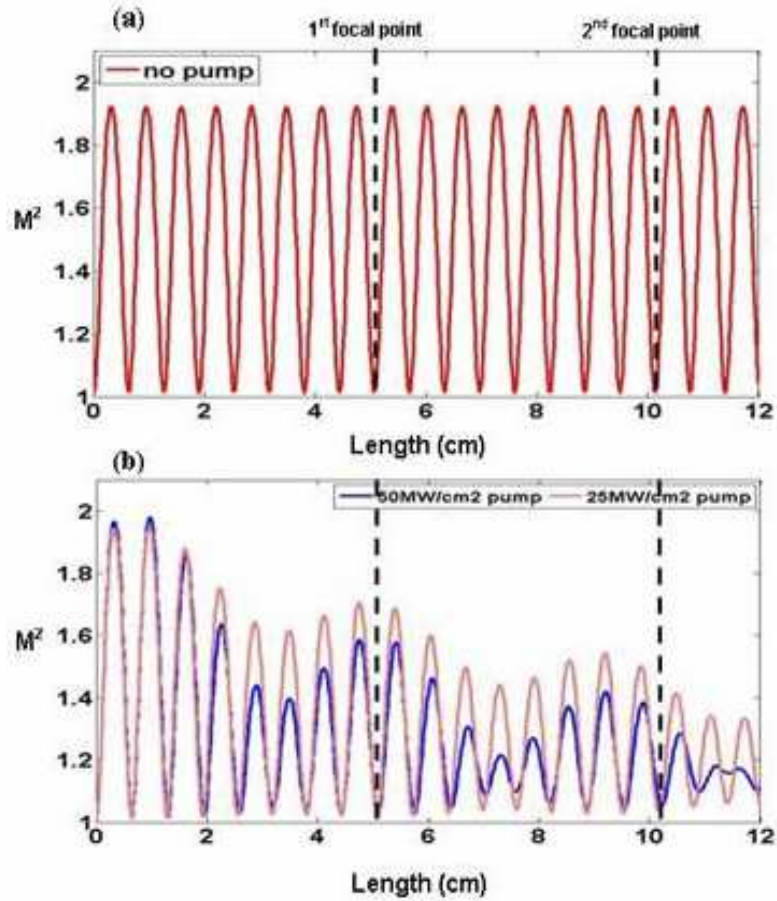

Fig. 4. The beam quality ($M^2$) of the Stokes beam calculated along the waveguide. The image periodically repeats itself due to the self-imaging property of the multimode waveguide. (a) Stokes beam propagating through a passive waveguide with no pump launched and (b) Stokes beam propagating through an active waveguide Raman amplifier with input pump intensities of 25MW/cm$^2$ and 50MW/cm$^2$. Pump powers are 1KW peak (1mW average) and 2KW peak (2mW average). The $M^2$ parameter is a measure of the image quality at the focal points with an ideal value of unity. The image deteriorates ($M^2$ increases) with increase in pump power due to preferential amplification of fundamental Stokes mode in comparison with higher order modes, and the presence of the phase-sensitive Raman four wave mixing.

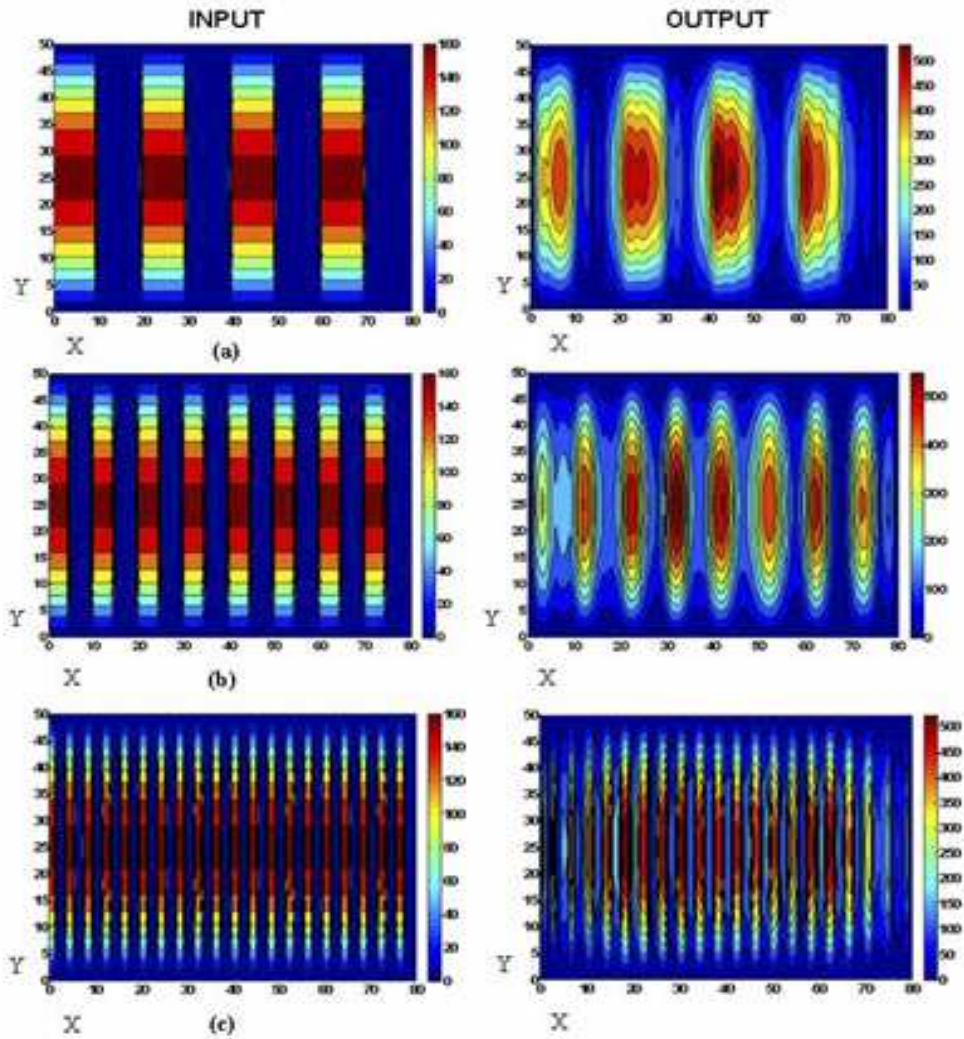

Fig. 5. The electric-field amplitude profile of a test image (left) and the amplified image (right). These figures describe the cross sectional (X-Y) profile as opposed to Figures 2 and 3 which show the propagation (X-Z) along the waveguide length. The spatial frequency of the test image is (a) 50 lines per mm, (b) 100 lines per mm, and (c) 250 lines per mm. Pump power of 1KW peak (1 mW average) is launched. The Stokes image experiences ~10dB gain over a length of ~5cm (ie. the first focal length).

## 4. Distortion-free image amplification using a single, high-order pump mode excitation

From the analysis of Raman amplification in a multimode waveguide with Gaussian-like transverse pump profile presented in sections 2 and 3 it is clear that there is a trade-off between the amount of image amplification and the distortion experienced by the image. However, there exists a special case in which distortion-free image amplification is possible. Since the pump is at a lower wavelength than the Stokes, the pump field supports more waveguide modes than the Stokes field. By propagating the pump only in a higher order mode which the Stokes image does not support, it is possible to eliminate the two sources of image distortion discussed in section 2. The RS-FWM terms vanish because the pump being in a single mode prevents the pump mode mixing effect (2$^{nd}$ term on RHS of Eqs. 5.1 and 5.2). Let $M$ be the index for the excited pump mode and $m<M$ be the Stokes mode index. With $k=M$, the self-coupling coefficient simplifies to:

$$\kappa_{mn-kl} = \kappa_{m1-M1} = \omega_S \varepsilon_o \chi_{Raman}^{(3)} \int_0^b \int_0^a |\phi_{P-M1}|^2 |\phi_{S-m1}|^2 dxdy = \omega_S \varepsilon_o \chi_{Raman}^{(3)} \cdot \frac{3 Z_P Z_S}{2ab} \quad (6)$$

which is independent of $m$, and hence is a constant for all Stokes modes. In this case the coupled mode equation for the Stokes mode simplifies to:

$$\frac{dA_{S-m1}}{dz} = A_{S-m1} \kappa_{m1-M1} |A_{P-M1}|^2, \; for\, all\, m<M \quad (7)$$

With a simple solution:

$$A_{S-m1}(z) = A_{S-m1}(0) \cdot \exp\left( \kappa_{m1-M1} \int_0^z |A_{P-M1}(z')|^2 dz' \right) \quad (8)$$

The exponential term represents the Raman gain experienced by the mode and is uniform for all the Stokes modes. Thus distortion free amplification is possible without trading off the gain. Selective mode excitation in multimode fibers is typically used for reducing the impact of modal dispersion in data communication applications [22]. Such techniques will prove powerful in utilizing the full potential of the silicon image amplifier.

## 5. Conclusions

In this paper we have proposed and analyzed a novel Raman amplifier in a multimode silicon waveguide. This amplifier consists of collinearly propagating pump and Stokes beams which are periodically self-imaged along the waveguide length. This ensures that the Stokes beam gets amplified and also reconstructs its profile at the focal points. We have also analyzed an application of this amplifier as an image pre-amplifier for MWIR remote sensing applications. The use of a multimode silicon waveguide as the active medium, with mature processing technology, large field-of-view and excellent thermal, damage and transmission properties in the MWIR lends itself to diverse image amplification applications. We have performed coupled-mode analysis of multimode waveguides including the conventional Raman terms and the phase-sensitive Raman four-wave mixing terms. We find that there are two different contributions to image distortion: (i) the conventional Raman terms can lead to preferential amplification of certain modes with the highest coupling coefficients at the expense of the other modes, and (ii) the RS-FWM terms which lead to phase distortions of the Stokes images and hence affect the self-imaging property. These introduce a tradeoff between the gain and image quality. One possible solution that can overcome this trade-off is to restrict the pump to a single high order spatial mode.

**Acknowledgements:**
This work was supported by DARPA and Northrop Grumman Corporation.